# Method Article

**Title:** A 'feasibility space' as a goal to be achieved in the development of new technologies for converting renewable energies

**Author:** Jones S. Silva[1], Fausto A. Canales[2], and Alexandre Beluco[1a]

**Affiliation:** 1 Instituto de Pesquisas Hidráulicas (IPH), Universidade Federal do Rio Grande do Sul (UFRGS), Porto Alegre, Rio Grande do Sul, Brazil;
2 Department of Civil and Environmental, Universidad de la Costa, Barranquilla, Atlantico, Colombia,

**Contact email:** albeluco@iph.ufrgs.br

**Abstract.** *This method article proposes the establishment of a feasibility space as an objective to be achieved during the development of new technologies to convert energy from renewable resources. The feasibility space can also be a reference when designing an energy system based on renewable resources. The feasibility space is a set of parameter values for the design stage that define the economic and technical feasibility of an energy system or a new technology, which must be satisfied when the energy system comes into operation or when the new technology for converting power goes into operation. The study of possible feasibility spaces allows characterizing energy systems or new technologies as attractive investments, or on the other hand, as unfeasible ventures.*
*- The method proposes to establish a goal to achieve during the development of technologies for energy conversion*
*- The method provides a benchmark for both the stages of design and development of generation systems and new technologies*
*- The feasibility space constitutes a planning tool for power systems based on renewable resources of any size*

**Keywords:** renewable energy, feasibility space, hybrid energy systems, new technologies.

**Graphical Abstract**

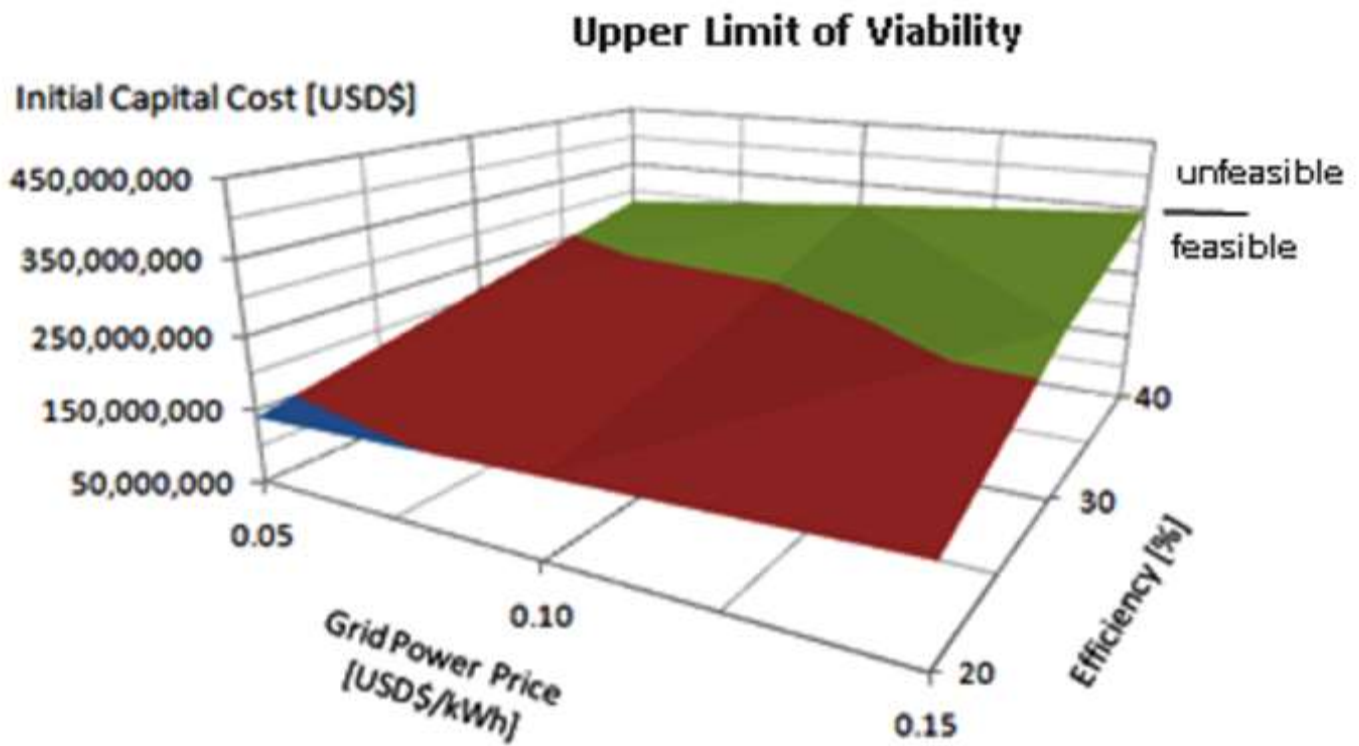

**SPECIFICATIONS TABLE**

| Subject Area | *Energy* |
|---|---|
| **More specific subject area:** | *Renewable Energy – Hybrid Energy Systems* |
| **Method name:** | *Feasibility space method, for determining the feasibility limits for the implementation of a generating plant based on renewable energy* |
| **Name and reference of the original method** | *Characterization of a feasibility space for a new technology – a case study of wave energy in southern Brazil. Silva et al. [1]. Current Alternative Energy (2018), v.2, n.2, p.112-122.* |
| **Resource availability** | *N/A* |

**Method details**

*Background*

Among the various renewable energies options, hydropower, solar energy, and wind energy have counted with mature and proven technology for their exploitation for some decades already. The exploitation of other renewable resources is still in developing stages, both technically and economically, for instance, wave energy and tidal energy. The work of Silva and Beluco [1] aimed at identifying attractive investments for the development of appropriate technology for the energy exploration of ocean waves on the northern coast of the State of Rio Grande do Sul, in southern Brazil. Previous studies [2] had indicated a potential for energy in ocean waves that can be considered attractive for investors from the energy sector. However, the attractiveness level of this investment depends on the characteristics of the energy system in which the devices for converting wave energy are connected. Based on this context, the concept of feasibility space [1] was developed. This space can change over time and depends on the energy system to which the hybrid system under study or the new technology under development is connected.

The feasibility space establishes a limit for parameters related to the project and which must be reached so that the investment in the energy system or in the technology under development is attractive. The feasibility space must be seen as an objective to be achieved with an adequate project for an energy system or with the development of a new technology, under penalty of not making the enterprise viable.

*Method*

The method for determining the feasibility space consists of the following five steps:

*1 - Establish the characteristics of the study to be carried out and the feasibility space to be configured, characterizing the hybrid energy system to be designed or the new technology to convert energy from renewable resources to be developed.*

*2 – Establish the parameters that can define the feasibility space of the energy system under analysis or the new technology under development and set the range of values for those parameters that may contain the feasibility limits.*
    *Note. Usually, these parameters can be the initial capital cost, the grid power price, the efficiency, among others.*

*3 - Perform simulations of the energy system under analysis or the new technology under development (with software such as Homer) to determine optimal solutions, within the intervals established for the feasibility analysis.*

*4 - If the results of the simulations indicate that some of the optimal solutions correspond to some value outside the ranges established in item 2 above, return to item 2 and reformulate the estimates for the ranges that define the feasibility space.*

*5 - Considering the final results of item 3 above, configure the feasibility space for the hybrid energy system to be designed or the new technology for converting energy from renewable resources to be developed.*

*Application example*

The original article [1] presenting the concept of feasibility space studied the investment viability for the development of devices for converting ocean wave energy on the southern coast of Brazil. The study carried out for the configuration of the feasibility space was based on simulations with the well-known [3] software Homer Legacy [4].

Following step 1 of the method described above, the work to be done in this case consists in determining a feasibility space for a new developing technology for converting ocean wave energy on the coasts of Southern Brazil. Following step 2, the variables that define the feasibility of the project, in this case, are the initial capital cost, the grid power price and the conversion efficiency.

The ranges of variation established for these variables, intending to delimit the feasibility space, are as follows: initial capital cost ranging between USD$ 50,000,000 and USD$ 450,000,000; grid power price ranging between USD$ 0.05 per kWh and USD$ 0.15 per kWh; and the efficiency ranging between 20% and 40%. These values appear as limits in the figure below.
The next step consists of simulations performed with the Homer Legacy software. This software simulates hybrid systems [5]-[6] over a year and selects as optimal solutions those with the lowest total net present cost over the analysis time, which is equal to 25 years in this case. The paper of Silva and Beluco [1] describes the simulations and shows how the feasibility space was built. The results of the simulations indicated that the ranges of variation of the variables, as established above, were adequate. The results were then used to configure the feasibility space, as shown in Table 1 and Figure 1. The simulated system is a PV-wind hybrid system operating near the coast and receiving energy from this hypothetical ocean wave power plant.

Table 1. Results delimiting the feasibility space for ocean wave power plants in southern Brazil.

| COEg | Eff. | Limit of viability | | COE |
|---|---|---|---|---|
| [USD$] | [%] | [USD$] | [USD$/kW] | [USD$] |
|  | 40 | 429,805,626.60 | 2,865.37 | 0.120 |
| 0.15 | 30 | 335,401,534.53 | 2,236.01 | 0.118 |
|  | 20 | 232,905,370.84 | 1,552.70 | 0.118 |
|  | 40 | 311,922,569.29 | 2,079.48 | 0.094 |
| 0.10 | 30 | 241,871,975.36 | 1,612.48 | 0.093 |
|  | 20 | 166,476,022.88 | 1,109.84 | 0.093 |
|  | 40 | 195,247,807.46 | 1,301.65 | 0.048 |
| 0.05 | 30 | 135,148,356,91 | 900.99 | 0.048 |
|  | 20 | 92,401,361,27 | 616.01 | 0.047 |

Legend: COEg is the cost of energy generated,
Eff is the efficiency, and COE is the cost of energy.

Table 1 indicates the results for the initial capital cost, in USD$, as a function of the cost of energy of the grid, in USD$ per kWh, and the projected efficiency for the wave energy conversion system. The values for the cost of energy from the grid and efficiency correspond to the intervals established above. The values obtained for the initial cost, as can be seen, also appeared within the range established above. These values establish an upper limit for the feasibility of investing in ocean wave energy in the studied area. The table also shows the investment amounts, in USD$, per installed kW.

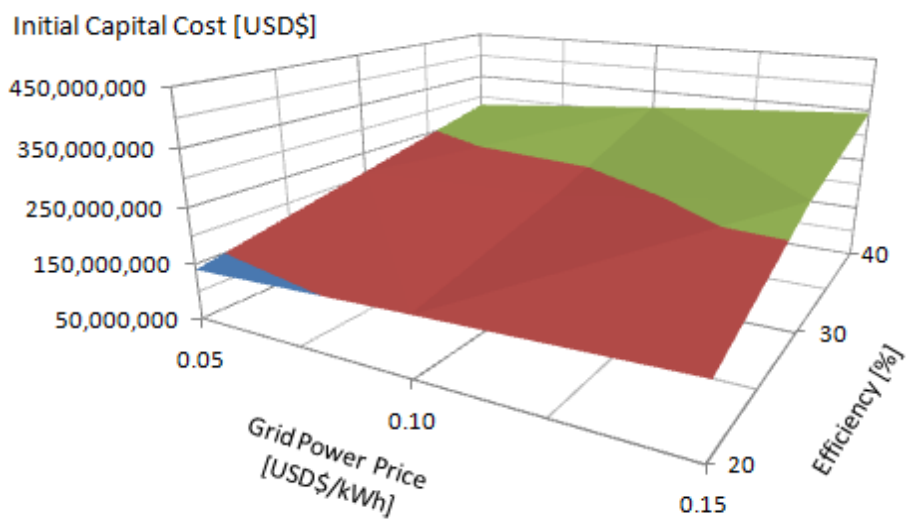

Figure 1. Upper limit of viability for the system of Ref. [1], considering the initial capital cost of the hybrid system as a function of grid power price and efficiency of the ocean wave power plant, with average wind speed equal to 7.62 m/s.

Figure 1 shows the limits that appear in Table 1, and they correspond to average wind speed in the case study region set as 7.62 m/s in the simulation. The three different colors used to fill the surface that appears in this figure correspond to three different bands for the initial capital cost. The highest values for the initial capital cost associated with the highest values for the grid power price and the highest efficiency values, as can be observed.

This surface shows how the feasibility can be improved with equipment that has better efficiency in converting ocean wave energy, but this improvement comes at the expense of a higher initial investment cost for the acquisition of energy conversion devices. This surface can vary its position over time, both by the evolution of technologies for converting wave energy and by the evolution of other technologies involved in the hybrid system that includes a connection to the wave energy power plant connected.

Thus, an investment in a plant for converting energy from ocean waves on the southern coast of Brazil will be viable if the initial capital cost of the project, as a function of the grid power price and the final efficiency of the developed equipment, is lower than the values establishing the viability limit, as established in Table 1 and Figure 1. Thus, the viability space is an objective, a goal, to be achieved.

**Acknowledgments:** *This work was developed as a part of research activities on renewable energy developed at the Instituto de Pesquisas Hidráulicas (IPH), Universidade Federal do Rio Grande do Sul (UFRGS). The authors acknowledge the support received by the institution. The last author acknowledges the financial support received from CNPq for his research work (proc. n.312941/2017-0).*